\documentclass[12pt,twoside]{article}
\usepackage{correl11}

\def\TheTitle{Hedin Equations, $GW$, $GW$+DMFT,\\ and All That}
\def\TheHeading{Hedin Equations, $GW$, $GW$+DMFT and All That}                        
\def\TheAuthor{K. Held,\\  C. Taranto, G. Rohringer, and  A. Toschi}                             
\def\TheInstitute{Institute for Solid State Physics}  
\def\TheAddress{Vienna University of Technology,  1040 Vienna, Austria}         
\def\TheChapter{5}                                       

\begin{document}
\MakeTitle           
\def\TheAuthor{Held {\em et al.}}               \tableofcontents     
\newpage

\section{Introduction}
\begin{figure}[tb]
 \centering
 \includegraphics[width=0.12\textwidth,angle=270]{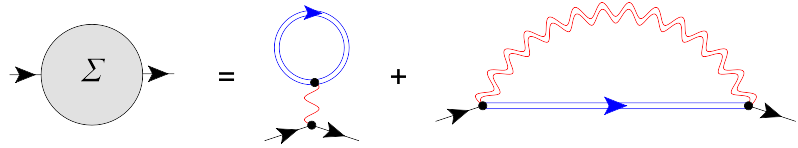}
 \caption{In $GW$, the self energy is given by the Hartree term plus a
 Fock-like term which is however in terms of the screened Coulomb interaction $W$ (double wiggled line) instead of the bare  Coulomb interaction $V$ (single wiggled line). The interacting Green function $G$ is denoted by a double straight line.\label{Fig:GWself}}
\end{figure}

An alternative to density functional theory \cite{Hohenberg64a,Jones89a,DFT} for
calculating materials {\em ab initio} is the so-called $GW$ approach \cite{Hedin,GWReview}.
The name stems from the way the self energy is calculated in this approach: It is given by the product of the Green function $G$ and the screened Coulomb interaction $W$,  see Fig.\ \ref{Fig:GWself}:
\begin{equation}
\Sigma^{\rm GW}({\mathbf r}, {\mathbf r^{\prime}};\omega) = {i} \int \frac{d \omega^{\prime}}{{2 \pi}}
 G({\mathbf r}, {\mathbf r^{\prime}};\omega+\omega^{\prime}) W({\mathbf r}, {\mathbf r^{\prime}};\omega^{\prime}) \; .
\label{SigmaGW}
\end{equation}
This self energy contribution supplements  the standard Hartree term (first diagram in Fig.\ \ref{Fig:GWself}.)
Here, ${\mathbf r}$ and  ${\mathbf r^{\prime}}$ denote two positions in real space, $\omega$ the frequency of interest; and the imaginary unit $i$ in front of
$G$ stems from the standard definition of the real time (or real frequency) Green function and the 
rules for evaluating the diagram  Fig.\ \ref{Fig:GWself}, see e.g.\ \cite{Abrikosov63}.

\begin{figure}[tb]
 \centering
 \includegraphics[width=0.18\textwidth,angle=270]{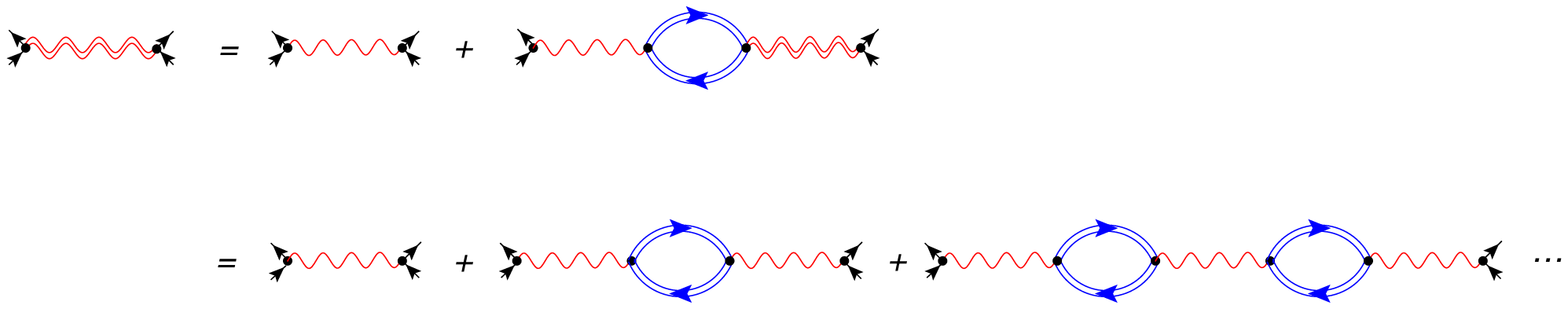}
 \caption{The screened interaction  $W$ (double wiggled line) is calculated from the bare Coulomb interaction $V$ (single wiggled line) and  corrections which describe screening processes. In $GW$, the screening is given by the random phase approximation, i.e., only bubble diagrams in a geometric series (second line) are considered. This geometric series can be generated from  a single bubble
connected to the screened interaction (first line). If we start with $W=V$ on the right hand side of the first line, we will generate the second term of the second line, and by further iterations  obtain the whole
series (whole second line).
\label{Fig:W}\label{Fig:RPA}}
\end{figure}

 In $GW$, the screened Coulomb interaction $W$ is calculated within the random phase approximation (RPA) \cite{Aryasetiawan}. That is, the screening is  given by the (repeated) 
interaction with independent electron-hole pairs, see Fig.\ \ref{Fig:W}.
For example, the physical  interpretation of the
second diagram in the second line of Fig.\ \ref{Fig:W}  is as follows: two electrons do not interact directly with each other as in the first term (bare Coulomb interaction) 
but the interaction is mediated via a virtual electron-hole pair (Green function bubble). Also included are repeated screening processes of this kind (third term etc.).
Because of the virtual particle-hole pair(s), charge is redistributed dynamically, and the electrons only see the other electrons through a  screened interaction. Since this RPA-screening is a very important contribution from the physical point of view and since, at the same time,  it is difficult to go beyond, one often restricts oneself to this approximation when considering screening, as it is done in $GW$.

\begin{figure}[tb]
 \centering
 \includegraphics[width=0.25\textwidth,angle=270]{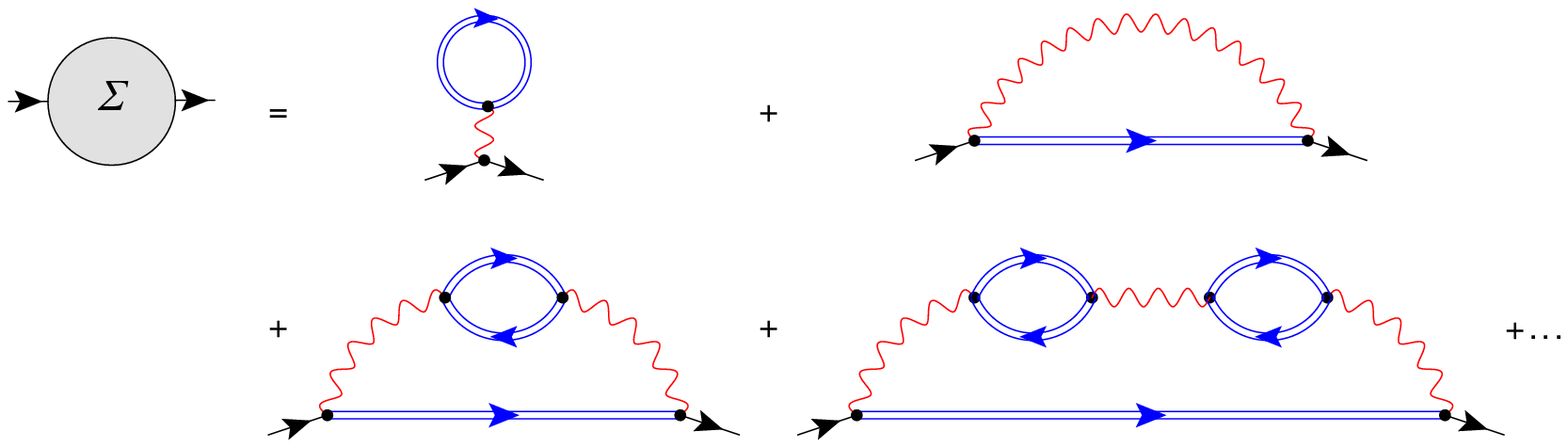}
 \caption{Substituting the screened interaction of Fig.\ \ref{Fig:W} into
the  self energy diagram  Fig.\ \ref{Fig:GWself} generates besides
the Hartree term (which is independently considered) the Fock term (first line), and some electronic correlations beyond (second line).
\label{Fig:GWcorr}}
\end{figure}

For visualizing what kind of electronic correlations are included in the $GW$
approach, we can replace the screened interaction in Fig.\ \ref{Fig:GWself}
by its RPA expansion
Fig.\ \ref{Fig:W}. This generates the Feynman diagrams of Fig.\ \ref{Fig:GWcorr}.
We see that besides the Hartree and Fock term (first line), 
additional diagrams  emerge (second line). By the definition that correlations are {\em what goes beyond Hartree-Fock}, these diagrams constitute
the electronic correlations of the $GW$ approximation. They give rise to 
quasiparticle renormalizations and finite quasiparticle lifetimes as well
as to renormalizations of the gaps in band insulators or semiconductors.

It is quite obvious that a restriction of the electronic correlations to 
only  the second line of  Fig.\ \ref{Fig:GWcorr} is not sufficient if electronic
correlations are truly strong such as in transition metal oxides or $f$-electron systems. The $GW$ approximation cannot describe Hubbard side bands or
Mott-Hubbard metal-insulator transitions. Its strength is for weakly 
correlated electron systems and, in particular, for
semiconductors. For these, extended sp$^3$ orbitals
lead to an important contribution of the non-local  exchange.
This is not well included (underrated) in the local exchange-correlation potential  $V_{\rm xc}$  of the local density approximation (LDA) and overrated by the bare Fock term. The $GW$ exchange is ``in between'' in magnitude and energy dependence, which is also important, as we will see later.
If it is, instead, the local correlation-part of  $V_{\rm xc}$ which needs to be
improved upon, as in the aforementioned  transition metal oxides or $f$-electron systems,
we need to employ dynamical mean field theory (DMFT) \cite{Metzner89a,Georges96a,DMFT} or similar many-body approaches.

Historically, the $GW$ approach was put forward by Hedin \cite{Hedin} as the simplest approximation to  the so-called Hedin equations.
In Section \ref{Sec:Hedin}, we will derive these Hedin equations from a Feynman-diagrammatical point of view. Section \ref{Sec:GW}  shows how $GW$ arises 
as an approximation to the Hedin equations.
In Section \ref{Sec:GWresults}, we will briefly present
some typical $GW$ results for materials, including the aforementioned
quasiparticle renormalizations, lifetimes, and band gap enhancements.
In Section \ref{Sec:GWDMFT}, the combination of $GW$ and DMFT is summarized.
Finally, as a prospective outlook, {\em ab initio} dynamical vertex approximation (D$\Gamma$A)
is introduced in  Section \ref{Sec:DGA} as a unifying scheme for all that: $GW$, DMFT and non-local vertex correlations beyond.

\section{Hedin equations}
\label{Sec:Hedin}
In his seminal paper \cite{Hedin}, Hedin noted, when deriving the equations
bearing his name: "The results [i.e., the Hedin equations] are well known to the Green function people''. And indeed, what is known
as the Hedin equations in the bandstructure community are simply the Heisenberg equation of motion for the self energy (also known as Schwinger-Dyson equation) and the
standard relations between irreducible and reducible vertex, self energy and Green function, polarization operator and screened interaction.
While Hedin gave an elementary derivation with only second quantization
as a prerequisite, we will discuss these from a Feynman diagrammatic point of view as the reader/student shall by now be familiar with this technique from the previous chapters/lectures of the Summer School.
Our point of view gives a complementary  perspective 
and sheds some light to the relation to standard many
body theory. For Hedin's     
elementary derivation based on functional derivatives
see \cite{Hedin} and \cite{GWReview}.

\begin{figure}[tb]
 \centering
 \includegraphics[width=0.185\textwidth,angle=270]{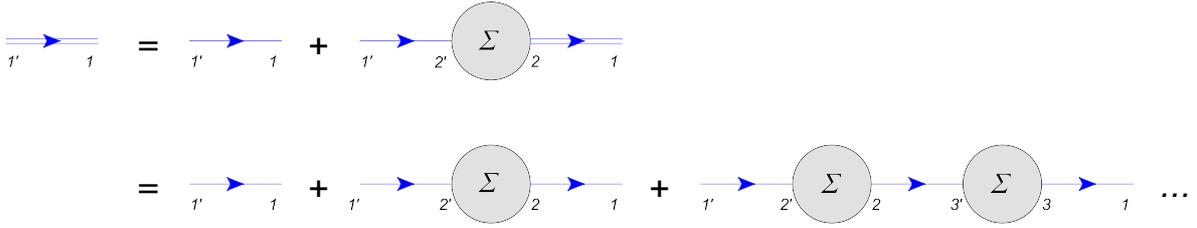}
 \caption{One of the five Hedin equations is the well-known 
Dyson equation, connecting the interacting Green function $G$  (double line),
non-interacting Green function $G^0$  (single line) and self energy $\Sigma$.
\label{Fig:Dyson}}
\end{figure}

Let us start   with the arguably simplest {\bf Hedin equation}: the well known
{\em Dyson equation}, Fig.\ \ref{Fig:Dyson}, which  connects self energy and Green function:
\begin{eqnarray}
G(11')&\!=\!&G^0(11')+G(12)\beta\Sigma(22')G^0(2'1')
\label{Eq:Dyson1} \\
&\!=\!&G^0(11')+G^0(12)\beta\Sigma(22')G^0(2'1') + G^0(13)\beta\Sigma(33')G^0(3'2)\beta\Sigma(22')G^0(2'1') + \ldots \nonumber
\end{eqnarray}
Here, we have introduced a short-hand notation with
$1$ representing a space-time coordinate $({\mathbf r_1},\tau_1)$ also subsuming a spin if present;
employ Einstein summation convention; 
$G$
and  $G^0$ denote the interacting and non-interacting ($V=0$) Green function\footnote{Please recall  the definition of the Green function with
Wick time-ordering operator {\cal T} for $\tau_1$ and $\tau_1'$:
\begin{eqnarray}
G(11') &\equiv & - \langle {\cal T} c(1) c(1')^{\dagger} \rangle \;.
\label{Eq:GF}\\
 &\equiv & -  \langle {\cal T} c(1) c(1')^{\dagger} \rangle \Theta(\tau_1-\tau_{1'}) +  \langle {\cal T} c(1')^{\dagger} c(1)  \rangle \Theta(\tau_{1'}-\tau_1) \label{Eq:GFdef}
\end{eqnarray}
The first term of Eq.\  (\ref{Eq:GFdef}) describes the propagation
of a particle from  $1'$ to $1$ (and the second line the corresponding propagation of a hole). Graphically we hence denote $G(11')$ by a straight line with an 
arrow from $1'$ to $1$ (The reader might note the reverse oder in $G(11')$; 
we usually apply operators from right to left).
},
respectively. Here and in the following part of this Section
we will consider the Green function in imaginary time with Wick rotation
 $t\rightarrow -i \tau$,
and closely follow the notation of \cite{Bickers}\footnote{Note that in  \cite{Bickers} summations are defined as $(1/\beta)\int_0^\beta{\rm d}\tau$ or in Matsubara frequencies $\sum_{m}$; Fourier transformations are defined as $G(\tau)=\sum_m e^{-i\nu_m\tau} G(\nu_m)$. The advantage of this definition is that the equations have then
the same form in $\tau$ and $\nu_m$. Hence, we employ this notation in this Section
and in Section 3. In Sections 1 and 4, the more standard definitions \cite{Abrikosov63} are employed, i.e.,  $\int_0^\beta{\rm d}\tau$; $(1/\beta)\sum_{m}$.
This results in some  factors $\beta$ (inverse temperatur), which can be ignored if one only wants to understand the equations.}.

In terms of Feynman diagrams, the Dyson equation means that we collect all one-particle {\em irreducible} diagrams, i.e., all diagrams that do not fall apart into two pieces if one Green function line is cut, and call this object $\Sigma$. All Feynman diagrams for
the interacting Green function are then generated simply by connecting the 
one-particle irreducible building blocks $\Sigma$ by Green function lines in the Dyson equation
(second line of  Fig.\ \ref{Fig:Dyson}). This way, no diagram is counted twice
since all  additional diagrams generated by the Dyson equation are one-particle
{\em reducible} and hence not taken into account for a second time.
 On the other hand all diagrams are generated:
the irreducible ones are already contained in $\Sigma$ and the
reducible ones have by definition the form of the second line of  Fig.\ \ref{Fig:Dyson}.

Let us note that the Dyson
equation (\ref{Eq:Dyson1}) can be resolved for
\begin{equation}
G(11')= \left( \left[ \left( G^0\right)^{-1} -\beta\Sigma \right ]^{-1} \right)_{(11')} 
\end{equation} 
with a matrix inversion in the spatial and temporal coordinates. It
is invariant under a basis transformation
from ${\mathbf r_1}$ to, say, an orbital basis or from time to frequency (note in a momentum-frequency basis, the $G$ and $\Sigma$ matrices are diagonal).

\begin{figure}[tb]
 \centering
 \includegraphics[width=0.28\textwidth,angle=270]{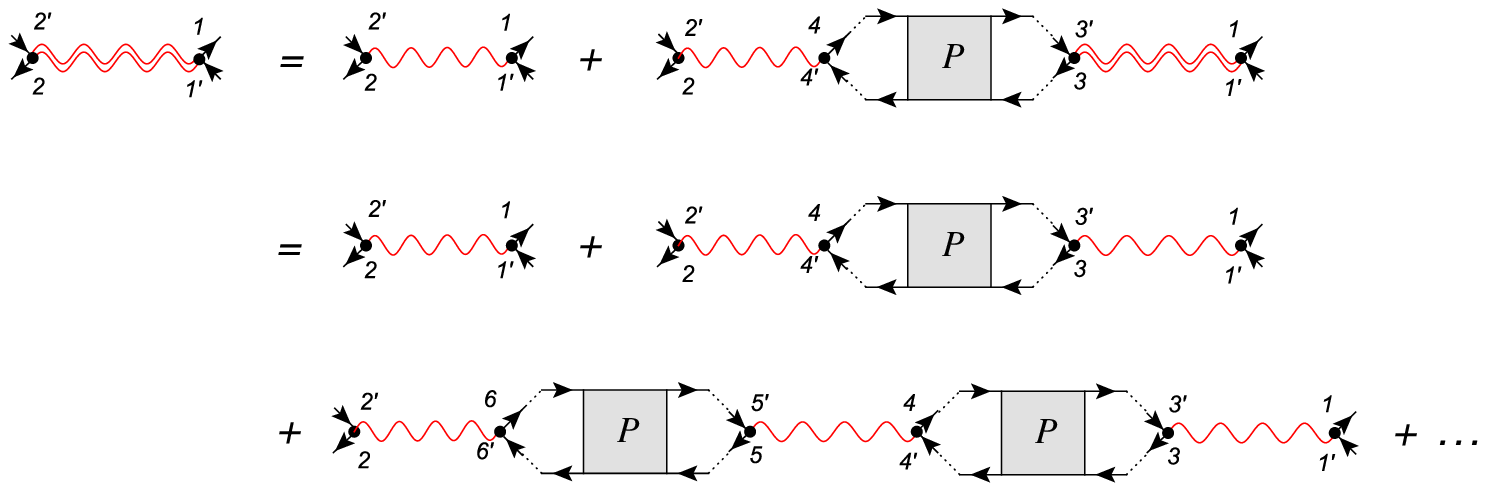}
 \caption{A second of the five Hedin equations is an ananlogon of the
Dyson equation, Fig.\ \ref{Fig:Dyson} but for the Coulomb interaction. 
It relates 
the screened Coulomb interaction $W$ (double wiggled line)
with the bare  Coulomb interaction $V$ and the polarization operator $P$.
As the Dyson equation can be considered as the defining equation for $\Sigma$,
this second Hedin equation effectively defines what $P$ is.
\label{Fig:WP}}
\end{figure}

A  second equation of the five Hedin equations actually has the same form as the {\em Dyson equation} but with the Green function and interaction changing their
role. It relates the {\em  screened Coulomb interaction $W$}
to the polarization operator $P$, see Fig.\ \ref{Fig:WP},
which is a generalization of  Fig.\ \ref{Fig:W} to arbitrary polarizations.
As $W$ we simply define (sum) all Feynman diagrams which connect to
the left and right side by interactions $V$.
Physically, this means that we  consider, besides 
the bare interaction, also all more complicated processes involving
additional electrons (screening). 

Similar as for the Dyson equation  (\ref{Eq:Dyson1}),
we collect all Feynman diagrams which do not fall apart into
a {\em left} and a {\em right} side by cutting one {\em interaction line $V$},
and call this object $P$. From $P$, we can generate all 
diagrams of $W$ connecting left and right side by a geometric
series (second and third line of Fig.\ \ref{Fig:WP}) with a repeated
application of $P$ and $V$ (second line of Fig.\ \ref{Fig:WP}).
As for the Dyson equation,
 we generate all Feynman diagrams (in this case for $W$) and count none twice this way.

Mathematically, Fig.\ \ref{Fig:WP} translates into a {\bf  second Hedin equation}
\begin{equation}
W(11';22')= V(11';22') +  W(11';33') P(3'3;4'4) V(44';22') \; .
\label{Eq:W}
\end{equation}
Note, that in a general basis the two particle objects have
four indices: an incoming particle 2' and hole 2, and  an  outgoing particle 1' and hole 1,
with possible four different orbital indices. In real space
two of the  indices are identical
 ${\mathbf r_1}={\mathbf r_{1'}}$ and  ${\mathbf r_2}={\mathbf r_{2'}}$.\footnote{\label{fV} This is obvious for the bare Coulomb interaction 
$V({\mathbf r_1},{\mathbf r_1'};{\mathbf r_2},{\mathbf r_2'})=V({\mathbf r_1},{\mathbf r_2})\delta({\mathbf r_1}-{\mathbf r_1'})\delta({\mathbf r_2}-{\mathbf r_2'})$ with
$V({\mathbf r_1},{\mathbf r_2})=\frac{e^2}{4\pi \epsilon_0}\frac{1}{|{\mathbf r_1}-{\mathbf r_2}|}$. As one can see in  Fig.\ \ref{Fig:WP} this property is
transfered to $W$ for which hence only a polarization with 
two  ${\mathbf r}$'s needs to be calculated in real space.}

\begin{figure}[tb]
 \centering
 \includegraphics[width=0.3\textwidth,angle=270]{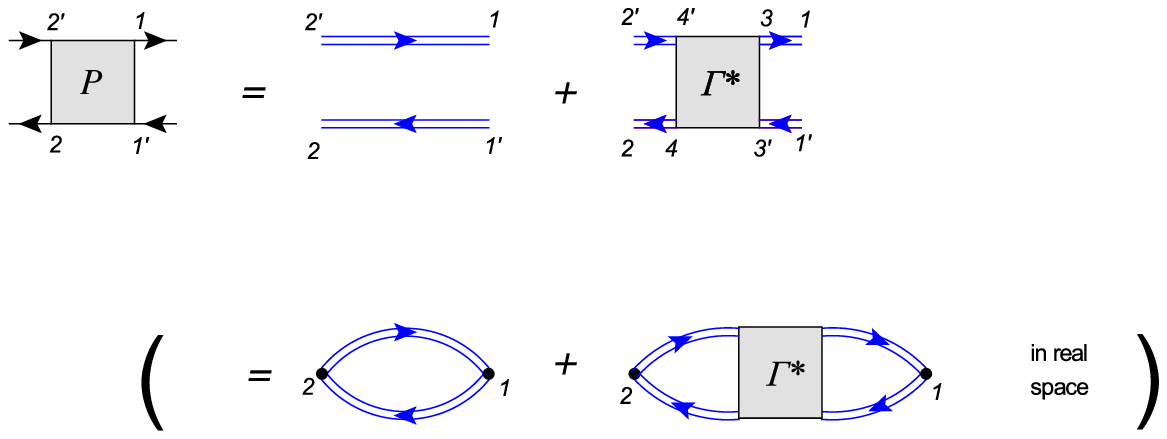}
 \caption{A third Hedin equation relates the polarization operator $P$ to  two separated  Green functions  (``bubble'' term)  plus vertex ($\Gamma^*$) corrections.
 This is the standard relation between two particle Green functions and  fully reducible vertex $\Gamma$. However since the polarization operator cannot include interaction-reducible diagrams  $\Gamma$ has to be replaced by $\Gamma^*$ (see text).
Second line: In terms of real space or momentum (but not in an orbital representation) two indices 
can be contracted to a single one (cf. footnote \protect \ref{fV}).
\label{Fig:P}}
\end{figure}

Next, we turn to the polarization operator $P$ which can be related to
a  vertex $\Gamma^*$. This is the standard relation between 
two particle Green functions (or response functions) and the vertex and
represents a {\bf third Hedin equation}. 
That is $P$ is given by the simple connection of left and right side by two (separated) Green functions plus vertex corrections, see  Fig.\ \ref{Fig:P}:
\begin{equation}
P(11';22')= \beta G(12') G(21') + \beta G(13) G(3'1')  \Gamma^{*}(33';44') \beta G(4'2') G(24) \; .
\label{Eq:HedinP}
\end{equation}
Note,  in real space $2=2'$ and $1=1'$ (cf. footnote \protect \ref{fV}) so that 
working with a two index object, see second line of Fig.\ \ref{Fig:P}, is possible (and was done by Hedin), the inverse temperature $\beta$ ($k_B\equiv 1$)
arises from the rules for Feynman diagrams in imaginary times/frequencies, see \cite{Bickers}.

\begin{figure}[tb]
 \centering
 \includegraphics[width=0.3\textwidth,angle=270]{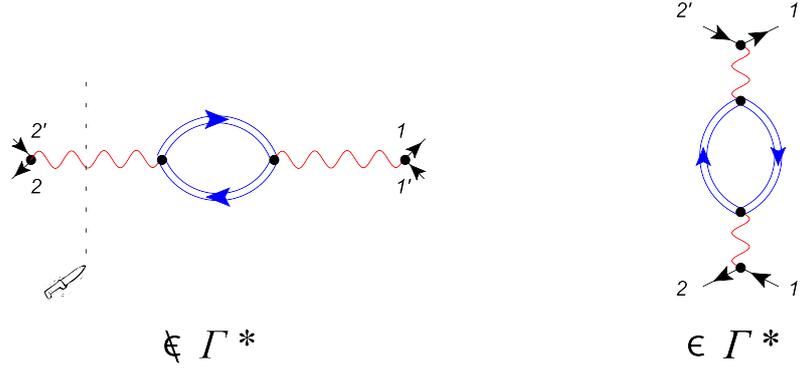}
 \caption{Left: A Feynman diagram that is {\bf not} part of  $\Gamma^*$ (or $P$)
since cutting a single interaction line
separates the diagram into left and right part. Right: This Feynman diagram is
included in  $\Gamma^*$ (or $P$).
\label{Fig:FDcut}}
\end{figure}

Let us keep in mind, that in $P$ or $\Gamma^*$ not all
Feynman diagrams are included: those diagrams, that  can  be separated into left and right
by cutting a single interaction line have to be explicitly excluded, see Fig.\ \ref{Fig:FDcut}. This is the reason why we put the symbol $^*$
to the vertex $\Gamma^*$; indicating that some diagrams of the full vertex $\Gamma$ are missing.

\begin{figure}[tb]
 \centering
 \includegraphics[width=0.17\textwidth,angle=270]{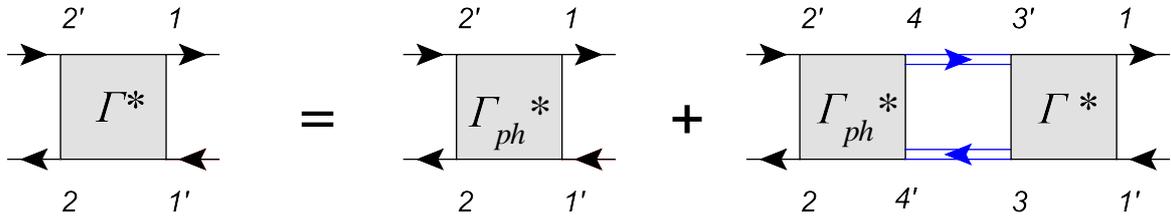}
 \caption{A fourth Hedin equation is the Bethe-Salpeter equation
between the irreducible $\Gamma^*_{\rm ph}$ and particle-hole reducible vertex $\Gamma^*$.
The particle-hole irreducible vertex  $\Gamma^*_{\rm ph}$ collects the  Feynman diagrams which cannot be
separated into left and right part
by cutting two Green function lines. All diagrams are then generated by the Bethe-Salpeter equation.
\label{Fig:Pirr}\label{Fig:BS}}
\end{figure}

Having introduced the two-particle vertex  $\Gamma^*$, the
{\bf fourth Hedin equation} is obtained by relating this vertex to another 
object, the particle-hole irreducible vertex  $\Gamma^*_{\rm ph}$. This relation is 
the standard Bethe-Salpeter equation.
As in the Dyson equation, we define the vertex $\Gamma^*$ as
the irreducible vertex   $\Gamma^*_{\rm ph}$ plus a geometric series
of repetitions of   $\Gamma^*_{\rm ph}$ connected by two Green functions, see Fig.\ \ref{Fig:BS}:
\begin{equation}
\Gamma^*(11';22')=\Gamma^*_{\rm ph}(11';22') + \Gamma^*(11';33') \beta G(3'4) G(4'3) \Gamma^*_{\rm ph}(44';22')
\label{Eq:BS}
\end{equation}
Here, the particle-hole irreducible vertex collects all Feynman diagrams
which cannot be separated  by cutting {\em two}
Green function lines into a left and right (incoming and outgoing) part.
The Bethe-Salpeter equation then 
generates all vertex diagrams by connecting the irreducible building blocks with
two Green function lines, in analogy to the Dyson equation for the
one-particle irreducible vertex $\Sigma$.

\begin{figure}[tb]
 \centering
 \includegraphics[width=0.17\textwidth,angle=270]{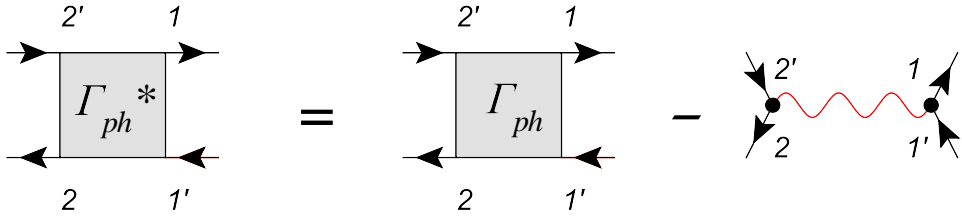}
 \caption{We have to exclude the bare Coulomb interaction $V$ from
the particle-hole vertex $\Gamma_{\rm ph}$ since such contributions are 
already considered in  Fig.\ \ref{Fig:WP}.
\label{Fig:ph}}
\end{figure}

Since in the polarization operator $P$ (and the corresponding reducible vertex vertex $\Gamma^*$) diagrams which connect
left and right by only one bare Coulomb interaction line $V$ are however excluded,
we have to explicitly take out this bare Coulomb interaction $V$ from the vertex $\Gamma^*$, i.e., we have the standard particle-hole vertex $\Gamma_{\rm ph}$ (i.e., all particle-hole irreducible diagrams) minus the bare Coulomb interaction $V$ diagram, see Fig.\ \ref{Fig:ph}:
\begin{equation}
\Gamma^*_{\rm ph}(11';22')=\Gamma_{\rm ph}(11';22')-V(11';22') 
\end{equation}
The bare $V$ and any combinations of $\Gamma^*_{\rm ph}$ and
$V$ are then generated in the screening equation (\ref{Eq:W}), Fig.\ \ref{Fig:WP}.

\begin{figure}[tb]
 \centering
 \includegraphics[width=0.55\textwidth,angle=270]{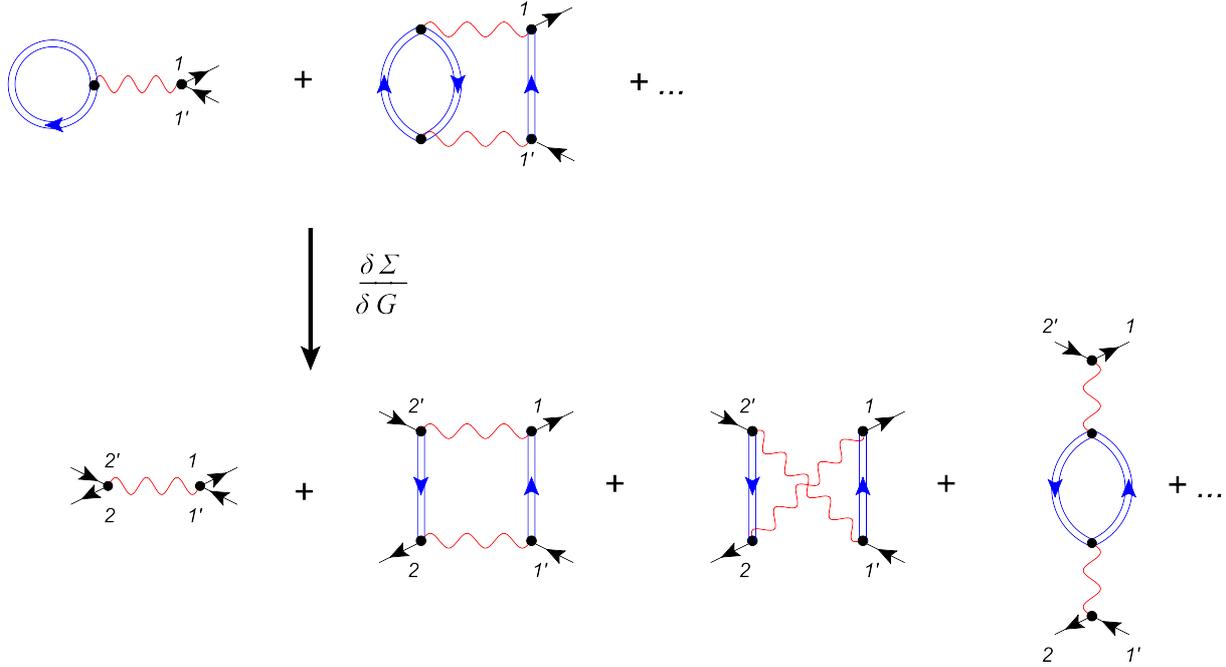}
 \caption{By hands of selected Feynman diagrams we illustrate that differentiation
of $\Sigma$ w.r.t. $G$ yields the particle-hole irreducible vertex.
\label{Fig:pSigma}}
\end{figure}

In this fourth  equation, Hedin directly expresses $\Gamma^*_{\rm ph}$ as the derivative of the self-energy w.r.t.\  the Green function \cite{GWReview} (respectively $V$ \cite{Hedin}).
This is a standard quantum field theoretical relation
\begin{equation}
\Gamma_{\rm ph}(11';22')=\frac{\delta \Sigma(11')}{\delta G(2'2)} \; ,
\label{Eq:Gammastardef}
\end{equation}
 which in terms of
Feynman diagrams follows from the observation that differentiation
w.r.t. $G$ means removing one Green function line, see Fig.\ \ref{Fig:pSigma} and  Ref.\ \cite{Bickers}. If we, as Hedin,  consider the self energy without Hartree term 
$\Sigma_{\rm Hartree}(11')=V(11';22')G(2'2)$ we obtain the
vertex $\Gamma^*_{\rm ph}$ instead of  $\Gamma_{\rm ph}$:
\begin{equation}
\Gamma^*_{\rm ph}(11';22')=\frac{\delta [\Sigma(11')-\Sigma_{\rm Hartree}(11')]}{\delta G(2'2)} 
\label{Eq:Gph}
\end{equation}
Note, the derivative of the  Hartree term w.r.t.\ $G(2'2)$ yields   $V(11';22')$, i.e., precisely 
the term  not included in  $\Gamma^*_{\rm ph}$, see first diagram of  Fig.\ \ref{Fig:pSigma}.

The {\bf fifth Hedin equation} is the {\em Heisenberg equation of motion} for the self energy, which follows
from the derivative of the Green function (\ref{Eq:GF}) w.r.t.  $\tau_1$, i.e.,
the time-part of the  coordinate $1$. Let us start with 
the Heisenberg equation
for the Heisenberg operator $c(1)$ ($\hbar\equiv 1$):
\begin{equation}
- \frac{\partial c({\mathbf r_1},\tau_1)}{\partial \tau_1} = [c({\mathbf r_1},\tau_1),H]
\label{Eq:HeisenbergC}
\end{equation}
and a general Hamiltonian of the form
\begin{equation}
H=H_0(11')  c(1)^{\dagger} c(1')+\frac{1}{2} V(11';22') c(1)^{\dagger} c(1') c(2)^{\dagger} c(2').
\end{equation}
From Eq.\ (\ref{Eq:HeisenbergC}), we obtain the Heisenberg equation of motion for the Green function:\footnote{
Note, the first term here is generated by the time derivative of the Wick
time ordering operator,    the second term stems from  $[c(1),H_0(22') c(2)^{\dagger} c(2')]$ and the third one from
  \[ 
[c(1), \frac{1}{2} V(33';22') c(3)^{\dagger} c(3') c(2)^{\dagger} c(2') ] \, \]
 employing 
$[A,BC] = B[A,C]+[A,B]C $ and the Fermi algebra
 $\{c(1'), c(1)^{\dagger}\}\equiv c(1')c(1)^{\dagger}+ c(1)^{\dagger}c(1')=\delta(1-1')$,  $\{c(1),c(1') \}=\{c(1)^{\dagger},c(1')^{\dagger}\}=0$.}
\begin{equation}
 -\frac{\partial G(11')}{\partial \tau_1} =    \delta(1-1')+H_0(12')G(2'1')-
V(13';22') \langle {\cal T} c(3') c(2)^{\dagger} c(2') c(1')^{\dagger} \rangle \; .
\label{Eq:HeisenbergGF}
\end{equation}

\begin{figure}[tb]
 \centering
 \includegraphics[width=0.4\textwidth,angle=270]{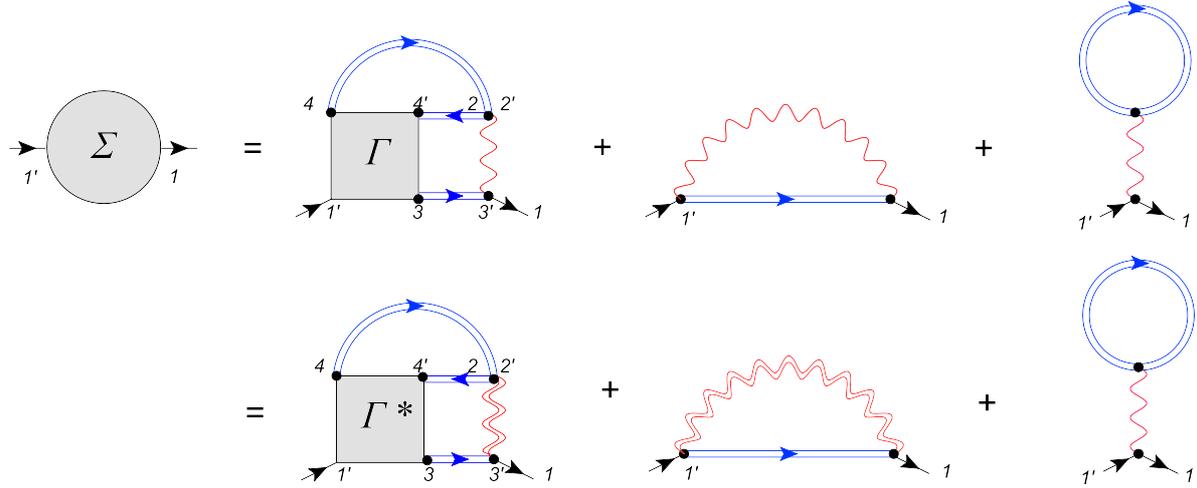}
 \caption{The fifth Hedin equation is the Heisenberg equation of motion,
which connects the one-particle Green function with the two particle
Green function or as shown in the figure (and as employed) the
self energy with the vertex.
\label{Fig:HeisenbergS}}
\end{figure}

The last term on the right hand side of Eq.\ (\ref{Eq:HeisenbergGF})
is by definition the self energy times the Green function.
Hence this combination
 equals the
interaction $V$ times a two-particle Green function.
The two particle Green function in turn is, in analogy to
Fig.\ \ref{Fig:P},  given by the bubble term (two Green function lines; there are actually two terms of this: crossed and non-crossed)
and vertex correction with the full (reducible) vertex $\Gamma$.
Besides the  Hartree  term, the self energy is given by (see Fig.\ \ref{Fig:HeisenbergS}):
\begin{equation}
\Sigma(11')=-V(13';22') \beta G(4'2) G(2'4) G(3'3) \Gamma(31';44')-V(12';21')G(2'2)
\label{Eq:SigmaVG}
\end{equation}
Note Eq. (\ref{Eq:SigmaVG}) can be formualted in an alternative way (see Appendix \ref{Sec:appendix} for a detailed calculation):
Instead of expressing this correlation part of the 
self energy by the bare interaction and the full vertex
we can take out the bare interaction line $V$ and any particle-hole repetitions
 of $V$ from the vertex, i.e.,
take $\Gamma^*$ instead of $\Gamma$.
Consequently we need to replace $V$ by $W$ to
generate the same set of all Feynman diagrams:
\begin{equation}
\Sigma(11')=-W(13';22') \beta G(4'2) G(2'4) G(3'3) \Gamma^*(31';44')-W(12';21')G(2'2)
\label {Eq:HeisenbergHedin}
\end{equation}
see second line of Fig.\ \ref{Fig:HeisenbergS}.\footnote{
Note, Hedin defines as a ``vertex'' $\Lambda$ \cite{GWReview} a combination
of $\Gamma^*$  and two Green function lines. Because  he works
in real space (or momentum) coordinates  only at 
a common point $2=2'$ needs to be considered (in the same way as in the second line of Fig.\ \ref{Fig:P}). Hedin  also adds a ``1'' in form of two $\delta$-functions:
\begin{equation}
\Lambda(11';2)=  \Gamma^*(11';33')  \beta G(3'2) G(32)+  \delta(1'-2) \delta(2-1)  \; .
\end{equation}
}

The five equations  
Eq.\ (\ref{Eq:Dyson1}), (\ref{Eq:W}),  (\ref{Eq:HedinP}),   (\ref{Eq:BS}), (\ref{Eq:HeisenbergHedin}) correspond to Hedin's equations (A13), (A20), (A24), (A22), (A23), respectively \cite{Hedin} [or to equations (44), (46), (38), (45), and (43), respectively, in \cite{GWReview}).
This set of equations is exact; it is equivalent
to the  text book quantum field theory \cite{Abrikosov63,Bickers} relations between
$\Gamma$,$\Gamma_{\rm ir}$, $\Sigma$ and $G$;  but it contains
additional equations since  $W$ and $P$ are introduced. 
The advantage is that, this way, one can develop much more directly 
approximations
where the screened Coulomb interaction plays a pronounced role
such as in the $GW$ approach.


\section{$GW$  approximation}
\subsection{From Hedin equations to $GW$}
\label{Sec:GW}
The simplest approximation is to neglect the vertex corrections
completely, i.e., to set set $\Gamma^*_{\rm ph}=0$.\footnote{Note this violates the Pauli principle, see last paragraph of Appendix \ref{Sec:appendix}.} Then  the Bethe-Salpeter equation (\ref{Eq:BS}) yields
\begin{equation}
 \Gamma^*=0 \;.
\end{equation}
The polarization in Eq.\ (\ref{Eq:HedinP}) simplifies to the bubble
\begin{equation}
P^{\rm GW}(11';22')=  \beta G(12') G(21') \;.
\label{Eq:PGW}
\end{equation}

The screened interaction in  Eq.\ (\ref{Eq:W})
 is calculated with this
simple polarization
\begin{equation}
W(11';22')= V(11';22') +  W(11';33') P^{\rm GW}(3'3;4'4) V(44';22') \; .
\end{equation}

The self energy in the Heisenberg equation of motion 
(\ref{Eq:HeisenbergHedin}) simplifies to Fig. \ref{Fig:W}, i.e.,
\begin{equation}
\Sigma^{\rm GW}(11')=-W(12';21')G(2'2)
\end{equation}
(plus Hartree term).

From this, the Green function is obtained via the Dyson equation (\ref{Eq:Dyson1}):
\begin{equation}
G(11')=G^0(11')+G(12)\beta\Sigma^{\rm GW}(22')G^0(2'1')
\end{equation}

These five (self-consistent) equations constitute the $GW$ approximation.

\subsection{$GW$ band gaps and quasiparticles}
\label{Sec:GWresults}

While the five $GW$ equations above are meant to be solved
self-consistently, most calculations hitherto started from a
LDA bandstructure calculation\footnote{An alternative, in particular for $f$ electron systems, is to use LDA+$U$ as a starting point \cite{Rinke}.}  and calculated from the LDA
polarization (or dielectric constant) a screened interaction
$W_0$ which in turn was used to determine the self energy
with the Green function $G_0$ from the LDA: 
$\Sigma=i G_0 W_0$.

Such calculations are already
pretty reliable for semiconductor band gaps, which
are  underestimated in the LDA. Due to the energy(frequency)-dependence
of $\Sigma$ bands at different energies are under the influence
of differently strong screened exchange contributions.
In semiconductors, it turns out that the conduction band is shifted
upwards in an approximately rigid way. The valence band is much less
affected so that the $GW$ band gap increases in comparison to the LDA gap.
This effect can be mimicked by a so-called scissors operator, defined
as cutting the density functional theory (DFT) bandstructure between valence and conduction
band and moving the conduction band upwards. Cutting LDA bandstructures
by a pair of scissors and rearranging them yields the $GW$ bandstructure
within an error of 0.1 eV for Si and 0.2 eV for GaAs \cite{Godby88}.

More recently, self-consistent $GW$ calculations became possible.
Many of these calculations employ an approximation of
 Schilfgaarde and Kotani \cite{P7:Faleev04,P7:Chantis06} where instead
of the frequency dependent $GW$ self energy $\Sigma_{nn'}(\omega,{\mathbf q})$
a frequency-independent Hermitian operator
\begin{equation}
\bar{\Sigma}_{nn'}={\rm Re} [\Sigma_{nn'}(\epsilon_q,{\mathbf q})+\Sigma_{n'n}(\epsilon_q,{\mathbf q})]/2
\label{Eq:SK}
\end{equation}
is constructed in the basis $n,n'$ employed in the $GW$/LDA algorithm. This self energy operator  has the
advantage that (as in LDA) we can remain in a one-particle description
and employ the  Kohn-Sham equations with Hermitian operator 
$\bar{\Sigma}_{nn'}$ 
to recalculate electron densities and Bloch eigenfunctions.

\begin{figure}
    \centering\includegraphics[height=7cm,clip=true]{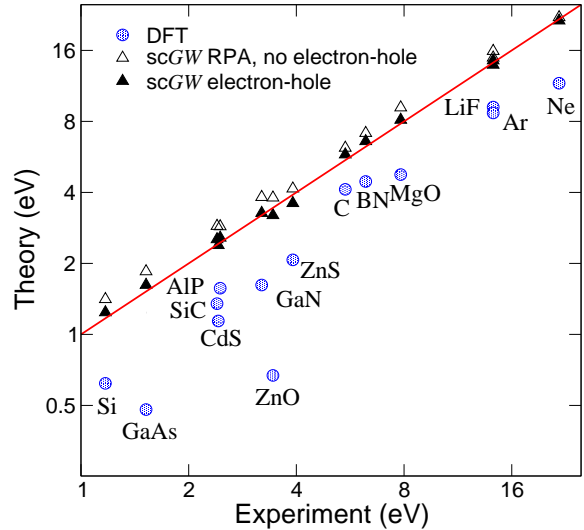}
    \caption{
Theoretical vs.\ experimental band gap of DFT, self-consistent $GW$ (sc$GW$) without and
with  (some) vertex corrections (electron-hole ladder diagrams) (reproduced from \cite{P7:Shishkin07}).\label{Fig:bandgaps}}
    \label{gK1}
  \end{figure}

The band gaps of this self-consistent approach are slightly larger
than experiment, see open triangles of Fig.\ \ref{Fig:bandgaps}.
This can be improved upon and band gaps can be calculated very
reliably if additional to $GW$ (some) vertex corrections are taken into
account. In  Fig.\ \ref{Fig:bandgaps}, the inclusion of electron-hole
ladder diagrams (visualized on the right hand side of Fig.\ \ref{Fig:bandgaps}) results in the
filled triangles with band gaps being within a few percent of the experimental
ones.
As in other areas of many-body theory, doing the self-consistency
without including vertex corrections does not seem to
be an improvement w.r.t. the non-self-consistent $GW$ since 
self-consistency and vertex corrections compensate each other in part. 
Full $GW$ calculations beyond the  Schilfgaarde and Kotani one-particle-ization
(\ref{Eq:SK}) have only been started and applied to simple
systems such as molecules \cite{Rostgaard,Stan}  and simple elements  \cite{Schoene,Kutepov}.

\begin{figure}
    \centering\includegraphics[height=8cm,clip=true]{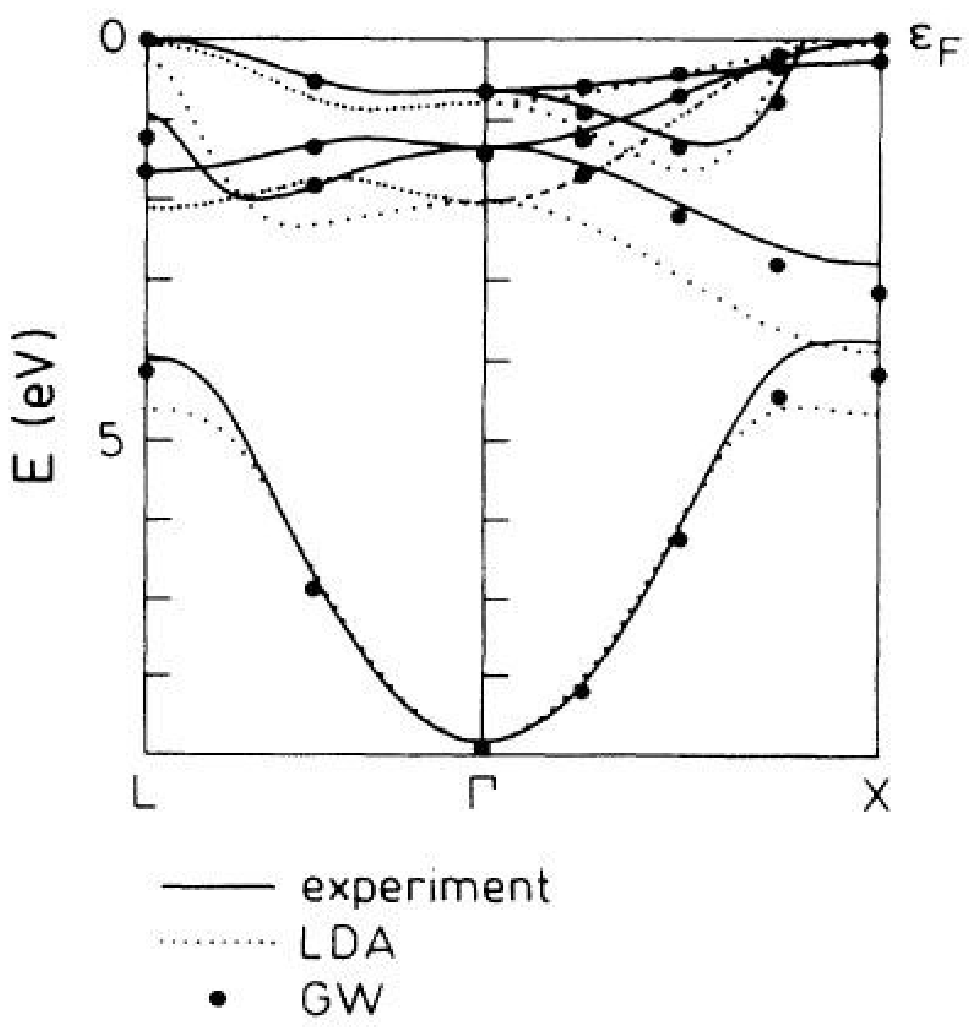}
    \caption{
Experimental, LDA and $GW$  bandstructure of Ni  (reproduced from \cite{Aritasetiawn92}).}
    \label{Fig:GWNi}
  \end{figure}

Besides this big success to overcome a severe LDA/DFT shortcoming for semiconductor gaps, $GW$ or
$G_0W_0$ calculations also show a quasiparticle renormalization of the bandwidth. For alkali metals, electronic correlation  are expected to be weak.
Nonetheless experiments observe e.g. in Na a band narrowing (of the occupied bands) of 0.6 eV \cite{Lyo88} compared to the nearly free electron theory.
While $GW$ \cite{Northrup87} yields such a band narrowing,  it
is quantitatively with 0.3 eV only half as large as in experiment \cite{Northrup87}.
Of the more strongly correlated transition metals, Ni is best studied:
here, the occupied $d$ bandwidth is 1.2 eV smaller in experiment than LDA 
and there is a famous satellite peak at -6eV in the spectrum
\cite{Huefner72}. While $GW$ \cite{Aritasetiawn92} yields a 
band-narrowing of 1 eV which is surprisingly good (see Fig.\ \ref{Fig:GWNi}),  the  satellite is
missing. In fact, it can be identified as a (lower) Hubbard band whose
description requires the inclusion of strong local correlation.
This is possible by DMFT; and indeed the satellite is
found in
LDA+DMFT 
\cite{FeNi}
and $GW$+DMFT calculations
\cite{Biermann03}, see next section.

Besides the mentioned band-narrowing which is associated with a reduced quasiparticle
weight  or effective mass enhancement (related to the real part of the self energy), there is also the imaginary part of the  $GW$  self energy,  which
corresponds to a scattering rate. For Ag the $GW$ scattering rate is
reported to be in close agreement with the experimental one 
obtained from two-photon photoemission 
\cite{Keyling}.

\section{$GW$+DMFT}
\label{Sec:GWDMFT}
Since $GW$ yields bandstructures similar to LDA (with the
 improvements for semiconductors  discussed in the previous section)
substituting the LDA part in LDA+DMFT by $GW$ is
very appealing from a theoretical point of view:
Both approaches $GW$ and DMFT are formulated in the
same many-body framework, which does not only has
the advantage of a more elegant combination, but also 
overcomes  two fundamental
problems of LDA+DMFT:
(i) The screened  Coulomb interaction employed for
$d$-$d$ or $f$-$f$ interactions in DMFT can be straight forwardly 
calculated via $W$; one does not need an additional constrained LDA approximation \cite{Dederichs84,McMahan88,Gunnarsson89}
to this end; (ii) the double counting problem, i.e.,
to subtract the LDA/DFT contribution 
of the local $d$-$d$ or $f$-$f$ interaction  which is 
included a second time in DMFT,
can be addressed in a rigorous manner since for $GW$+DMFT we actually know which Feynman diagram is counted twice.

\begin{figure}
    \centering\includegraphics[width=0.25 \textwidth,clip=true,angle=270]{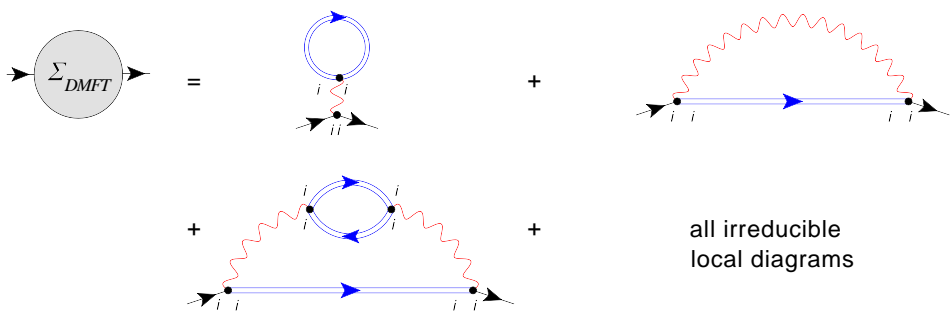}
    \caption{
The DMFT self energy is calculated from  the local contribution of all (one-particle irreducible) Feynman diagrams.
    \label{Fig:DMFT}}
  \end{figure}

 Biermann {\em et al.} \cite{Biermann03}  proposed  $GW$+DMFT, which they
discuss from 
a functional integral point of view: a $GW$ functional and a local impurity functional are added; the derivatives
yield the mixed $GW$+DMFT equations.
From a Feynman diagrammatic point of view, this corresponds to 
adding the $GW$ self energy, Fig.\ \ref{Fig:GWself} and the DMFT self energy which is just given
by the local contribution of all (one-particle irreducible) Feynman diagrams,
see Fig.\ \ref{Fig:DMFT}. From these, the  {\em local} screened exchange $GW$
and the doubly counted Hartree term need obviously to be explicitly subtracted for
not counting any diagram twice.

This results in the  algorithm Fig.\ \ref{flowGWDMFT}. Here, we leave the
 short-hand notation of Section \ref{Sec:Hedin} and \ref{Sec:GW} with $1$, $2$  since $GW$ is diagonal in $\omega$ and
 ${\mathbf k}$ and DMFT is diagonal in  $\omega$ and site indices.
Let us briefly discuss the $GW$+DMFT algorithm step-by-step; for more details see \cite{Held07}:

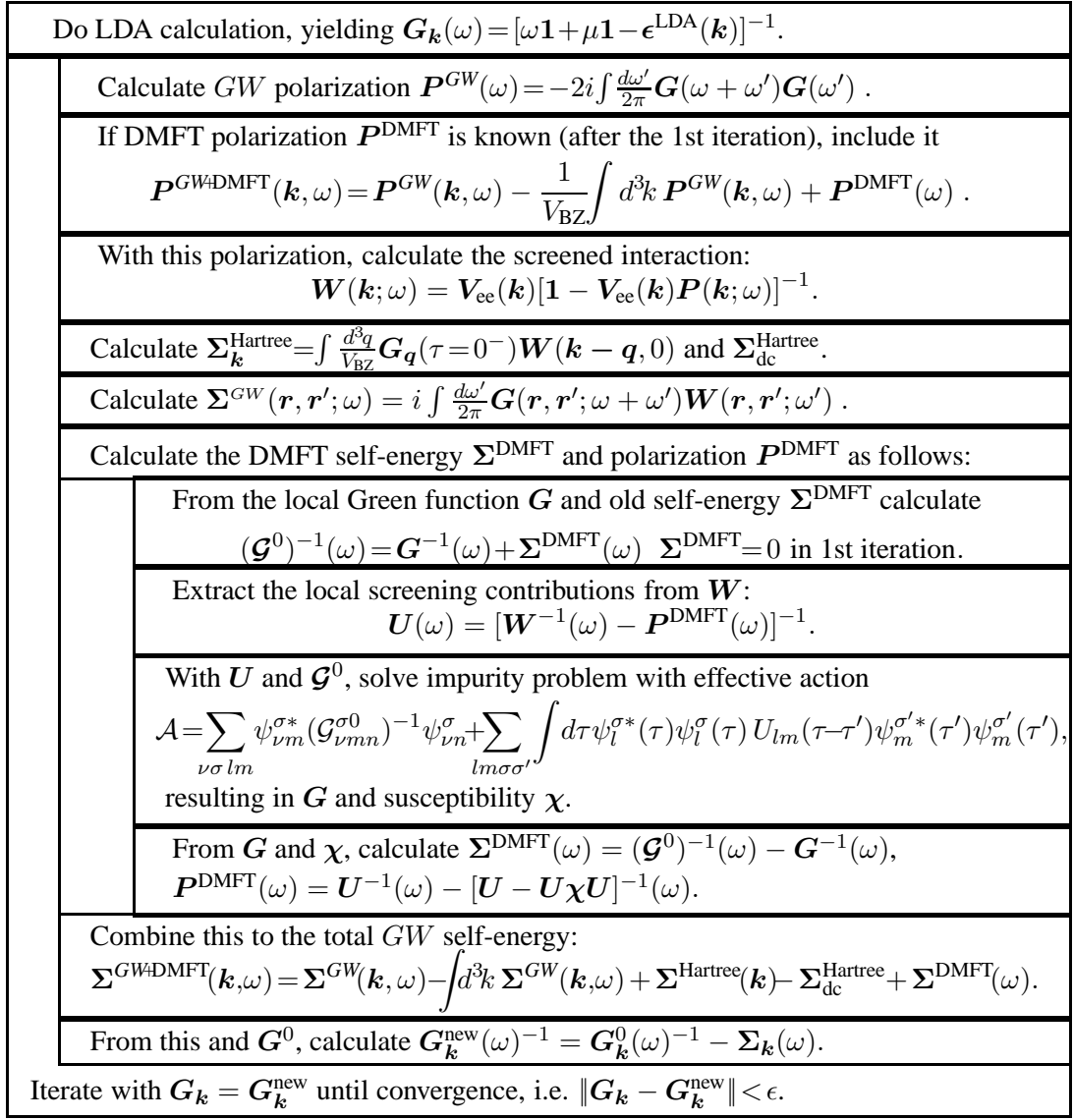
\begin{figure}[htb]
\begin{center}
\unitlength1mm
\small
\hspace{-.15cm}\begin{picture}(150,184)
\thicklines

\put(3.00,175.00){\framebox(142.00,7.00)[cc]
{\parbox{130mm}{Do  LDA calculation, yielding 
	$\bm{G}_{\bm{k}}(\omega)\!=\! [\omega\bm{1}\!+\!\mu\bm{1}\!-\! {\bm{\epsilon}}^{\text{LDA}}{(\bm{k})}]^{-1}$.
        }}}

\put(3.00,32.00){\framebox(142.00,143.00)[cc]{}}

\put(10.00,167.00){\framebox(135.00,8.00)[cc]
{\parbox{125mm}{ \vspace{.05cm}

Calculate $GW$  polarization 
$\bm{P}^{\text{{\em GW}}} (\omega) \!= \! - 2 i  \! \int \! \frac{d \omega^{\prime}}{{2 \pi}}
 \bm{G}(\omega+\omega^{\prime}) \bm{G}(\omega^{\prime})\;.
$
}}}

\put(10.00,151.00){\framebox(135.00,16.00)[cc]
{\parbox{125mm}{\vspace{.2cm}

If  DMFT polarization $\bm{P}^{\text{DMFT}}$  is known (after the 1st iteration), include it
\vspace{-.3cm}

\[
\! \!\bm{P}^{{\text{{\em GW}\!+\!DMFT}}} (\bm{k}, \omega)\! =\!\bm{P}^{\text{{\em GW}}} (\bm{k}, \omega)
-\frac{1}{V_{\text{BZ}}}\!{\int d^3\!k\, \bm{P}^{\text{{\em GW}}}(\bm{k}, \omega)}+
\bm{P}^{\text{DMFT}} (\omega) \;.
\]
}}}

\put(10.00, 139.00){\framebox(135.00,12.00)[cc]
{\parbox{125mm}{\vspace{.15cm}

With this polarization, calculate the screened interaction:
\vspace{-.5cm}

\[
 \bm{W}(\bm{k}; \omega)  =
\bm{V}_{\text{ee}}(\bm{k})
 [\bm{1}
 - \bm{V}_{\text{ee}}(\bm{k}) \bm{P}(\bm{k};\omega)]^{-1}.
\]}}}

\put(10.00, 132.00){\framebox(135.00,7.00)[cc]
{\parbox{127mm}{ Calculate  ${ \bm{\Sigma}}^{\text{Hartree}}_{\bm{k}}\!\!=\! \! \int  \frac{d^3\!q}{V_{\text{BZ}}} \bm{G}_{\bm{q}}(\tau\!=\!0^-)  \bm{W}(\bm{k-q},0)$ and
 $\bm{\Sigma}^{\text{Hartree}}_{\text{dc}}$.
               }}}

\put(10.00,125.00){\framebox(135.00,7.00)[cc]
{\parbox{127mm}{
 Calculate
$
 \bm{\Sigma}^{\text{$\scriptstyle GW$}}(\bm{r}, {\bm{r}^{\prime}};\omega) = {i} \int \frac{d \omega^{\prime}}{{2 \pi}}
 \bm{G}(\bm{r}, {\bm{r}^{\prime}};\omega+\omega^{\prime}) \bm{W}(\bm{r}, {\bm{r}^{\prime}};\omega^{\prime})
\; .$
             }}}

\put(10.00,118.00){\framebox(135.00,7.00)[cc]
{\parbox{127mm}{ \vspace{.1cm}

Calculate the DMFT self-energy ${\bm{\Sigma}}^{\text{DMFT}}$ and polarization $\bm{P}^{\text{DMFT}}$ as follows:}}}

\put(20.00,106.00){\framebox(125.00,12.00)[cc]
{\parbox{115mm}{ \vspace{.3cm}

From the local  Green function
 $\bm{G}$ and old self-energy ${\bm{\Sigma}}^{\text{DMFT}}$ 
calculate
\vspace{-.3cm}

   \[ 
(\mbox{\boldmath ${\cal G}$}^0)^{-1}(\omega)\!=\! \bm{G}^{-1}(\omega)\! + \! {\bm{\Sigma}}^{\text{DMFT}}(\omega)\;\; {\bm{\Sigma}}^{\text{DMFT}}\!\!=\!0 \,\, \mbox{in 1st iteration}.
\]   
}}}

\put(20.00, 94.00){\framebox(125.00,12.00)[cc]
{\parbox{115mm}{ \vspace{.2cm}

Extract the local screening contributions from $\bm{W}$:
\vspace{-.5cm}

   \[ \bm{U}(\omega)= [\bm{W}^{-1}(\omega) -  {\bm{P}}^{\text{DMFT}}(\omega)]^{-1}.
\]  

 }}}

\put(20.00, 71.00){\framebox(125.00,23.00)[cc]
{\parbox{117mm}{ \vspace{-.1cm}

With  $\bm{U}$  and $\mbox{\boldmath ${\cal G}$}^0$,
solve  impurity problem with effective action
\vspace{-.72cm}

\[
\!\! {\cal A}\!=\!\!\!\sum_{\nu
\sigma \,l m}\!\psi _{\nu
m}^{\sigma \ast }({\cal G}^{\sigma 0}_{\nu m n})^{-1}\psi _{\nu n}^{\sigma
{\phantom\ast }}
\!+\!\!\!\sum_{l m\sigma\sigma^{\prime}}
\!
    \int\limits\!d\tau\psi _{l}^{\sigma \ast }(\tau )
\psi_{l}^{\sigma}(\tau)
\,U_{l m}(\tau\!-\!\tau^{\prime})
\psi _{m}^{\sigma^{\prime} \ast }(\tau^{\prime} ) \psi_{m}^{\sigma^{\prime}}(\tau^{\prime} ),
\]\vspace{-.5cm}

resulting in  $\bm{G}$  and susceptibility $\bm{\chi}$.
              }}}

\put(20.00,59.00){\framebox(125.00,12.00)[cc]
{\parbox{115mm}{ \vspace{-.05cm}

From  $\bm{G}$ and  $\bm{\chi}$,
calculate 
   $  {\bm{\Sigma}}^{\text{DMFT}}(\omega)=  (\mbox{\boldmath ${\cal G}$}^0)^{-1}(\omega)-\bm{G}^{-1}(\omega)
$,
 \vspace{.05cm}

$ {\bm{P}}^{\text{DMFT}}(\omega) = \bm{U}^{-1}(\omega)-
[\bm{U}- \bm{U} {\bm{\chi}}\bm{U}]^{-1}(\omega).
$
 }}}
\put(10.00,45.00){\framebox(135.00,80.00)[cc]{}}

\put(10.00, 45.00){\framebox(135.00,14.00)[cc]
{\parbox{127mm}{\vspace{.45cm}

Combine this to the total $GW$ self-energy:
\vspace{-.85cm}

\[
{\bm{\Sigma}}^{{\text{{\em GW}\!+\!DMFT}}} \! (\bm{k}, \! \omega)\! =\!{\bm{\Sigma}}^{\text{{\em GW}}}\! (\bm{k}, \omega)
\!-  \! \! \!\int \! \!  \!d^3\!k\, {\bm{\Sigma}}^{\text{{\em GW}}} (\bm{k}, \! \omega) 
+ {\bm{\Sigma}}^{\text{Hartree}}  \! (\bm{k}) \! \!
- {\bm{\Sigma}}^{\text{Hartree}}_{\text{dc}}\!+{\bm{\Sigma}}^{{\text{DMFT}}} \! (\omega).
\]
}}}

\put(10.00, 39.00){\framebox(135.00,6.00)[cc]
{\parbox{127mm}{\vspace{.2cm}
From  this and   $\bm{G}^0$, calculate $ 
 \bm{G}^{\text{new}}_{\bm{k}}(\omega)^{-1} = 
\bm{G}^0_{\bm{k}}(\omega)^{-1} - {\bm{\Sigma}}_{\bm{k}} (\omega) .
$
\vspace{.15cm}

}}}

\put(5.00,34.00){
\parbox{115mm}{ Iterate with  $ \bm{G}_{\bm{k}}=\bm{G}^{\text{new}}_{\bm{k}}$ until
convergence, i.e. $|\!|\bm{G}_{\bm{k}}-\bm{G}^{\text{new}}_{\bm{k}}|\!|\! <\! \epsilon$.
}}
\end{picture}
\end{center}
\vspace{-3.4cm}

\caption{Flow diagram of the $GW$+DMFT algorithm (reproduced from \cite{Held07}).
\label{flowGWDMFT} \label{P7:Fig:GWflow}}
\end{figure}
\begin{itemize}
\item
In most $GW$ calculations, the starting point is 
a conventional LDA calculation (or another suitably chosen generalized Kohn-Sham calculation), 
yielding an electron 
density $\rho({\mathbf r})$, bandstructure  $\mbox{$\epsilon$}^{\rm LDA}{({\bf k})}$ and also 
an LDA Green function ${G}_{\mathbf k}(\omega)$ (the latter is calculated as
 in the first line of Fig.\ \ref{P7:Fig:GWflow}, where bold symbols denote an (orbital) 
matrix representation).

\item
From this Green function, the independent particle 
polarization  operator $P^{GW}$ is calculated  convoluting
two Green functions (2nd line of flow diagram   Fig.\ \ref{P7:Fig:GWflow}).
 Note there is a factor of 2 for the spin.

\item
From the polarization operator in turn, the local polarization has to be subtracted since this can (and has to)
be calculated more precisely within DMFT, which includes more than 
the RPA  bubble diagram (after the first DMFT iteration).

\item 
Next, the screened interaction $W$ is calculated from the bare Coulomb interaction $V_{\rm ee}$
and the overall polarization operator $P^{GW+DMFT}$.

\item Now, we are in the position to calculate the $GW$ self energy. The first term
is the Hartree diagram, which can be calculated straight forwardly in imaginary time $\tau$, yielding
$\Sigma^{\rm Hartree}$ and the corresponding local contribution $\Sigma^{\rm Hartree}_{\rm dc}$,
which we need to subtract later  to avoid a double counting as it is also contained in the DMFT.
\item 
The second diagram is the exchange from   Fig.\ \ref{Fig:GWself} which has the form
$G$ times $W$ for the $GW$ self energy.

\item
This $GW$ self energy has to be supplemented by the local DMFT self energy, which together
with the DMFT polarization operator is calculated in the following four steps:

\begin{enumerate}
\item The non-interacting  Green function ${\cal G}^0$ which defines a corresponding 
Anderson  impurity model is calculated.
\item The local (screened) Coulomb interaction $U(\omega)$ has to be determined {\em without} the local screening contribution,
 since the local screening will be again included in the DMFT. That is we have to ``unscreen''
$W$ for these contributions.
\item The Anderson impurity model defined by  ${\cal G}^0$ and  $U(\omega)$ has to be solved for its 
interacting Green function $G(\omega)$ and two-particle charge susceptibility $\chi$.
This is numerically certainly the most demanding step.
\item From this $G(\omega)$ and  ${\cal G}^0$, we obtain a new DMFT self energy $\Sigma(\omega)$ and 
from the  charge susceptibility a new DMFT polarization operator.
\end{enumerate}
\item
All three terms of the self-energy have now to be added; and the local screened exchange and Hartree contribution
need to be subtracted to avoid a double counting.
\item From this $GW$+DMFT self energy we can finally recalculate the Green function and iterate
 until convergence.

\end{itemize}

 The flow diagram already shows that the $GW$+DMFT approach is much more involved than LDA+DMFT.
However, it has the advantage that the double counting problem is solved and also the Coulomb interaction is calculated {\em ab initio} in a well defined and controlled way. Hence, no {\em ad hoc} formulas or parameters need to
be introduced or adjusted.

For defining a well defined interface between $GW$ and DMFT a particular problem is that
$GW$ is naturally formulated in real or $\mathbf k$ space and is presently implemented, e.g., in the 
LMTO \cite{GWReview} or  PAW basis \cite{P7:Shishkin06}.
However, on the DMFT side we do need to identify the interacting {\em local} $d$- or $f$-orbitals 
on the sites of the transition metal or rare earth/lanthanoid sites, respectively.
The switching between these two representations is non-trivial. It 
can be done
 by a downfolding  \cite{P7:Andersen99,P7:AndersenPsik} or a projection onto Wannier orbitals, e.g., using maximally localized Wannier orbitals 
\cite{P7:Marzari97,Wien2Wannier} or a simpler projection onto the  $d$ (or $f$) part of the wave function
within the atomic spheres  \cite{P7:Anisimov05,P7:Aichhorn09}. 
However, not only the one-particle wave functions and 
dispersion relation need to be projected onto the interacting subspace but also the interaction itself.
To approach the latter, a constrained random phase approximation (cRPA) method has been proposed  \cite{P7:Aryasetiawan04,Aryasetiawan} and improved by disentangling the $d$(or $f$)-bands \cite{P7:cRPA2}. The 
latter improvement now actually allows us to do cRPA in practice.
For the calculation of  the two-particle polarization operators and interactions,
Aryasetiawan {\em et al.} \cite{P7:AryasetiawanCP04} even proposed to use a third
 basis: the optimal product basis.

On the DMFT side, the biggest open challenges are to actually perform the DMFT calculations with 
a frequency dependent Coulomb interaction $U(\omega)$ and to calculate the DMFT charge susceptibility or polarization operator.

As the  fully self-consistent  $GW$+DMFT scheme is a formidable task,
Biermann {\em et al.}  \cite{Biermann03} employed a simplified implementation for their $GW$+DMFT 
calculation  of Ni, which is  actually the only $GW$+DMFT calculation hitherto:
For the DMFT impurity problem, only the local  Coulomb interaction between $d$ orbitals was included
and its frequency dependence was neglected ${\mathbf W}(\omega)\approx{\mathbf W}(0)$.
Moreover, only one iteration step has been done, calculating the
inter-site part of the self energy by $GW$ with the LDA Green function as an input
and the intra-site part of the self energy by DMFT (with the usual DMFT
self-consistency loop).
The $GW$ polarization operator ${\mathbf P}^{\rm GW}$ was calculated from the LDA instead of the $GW$ Green function.
This is, actually,  common practice even for  conventional $GW$ calculations
which are often of the $G_0W_0$ form (see Section \ref{Sec:GWresults}).

\begin{figure}[tb]
{\hspace{-0.5cm} \includegraphics[clip=true,width=7.099cm]{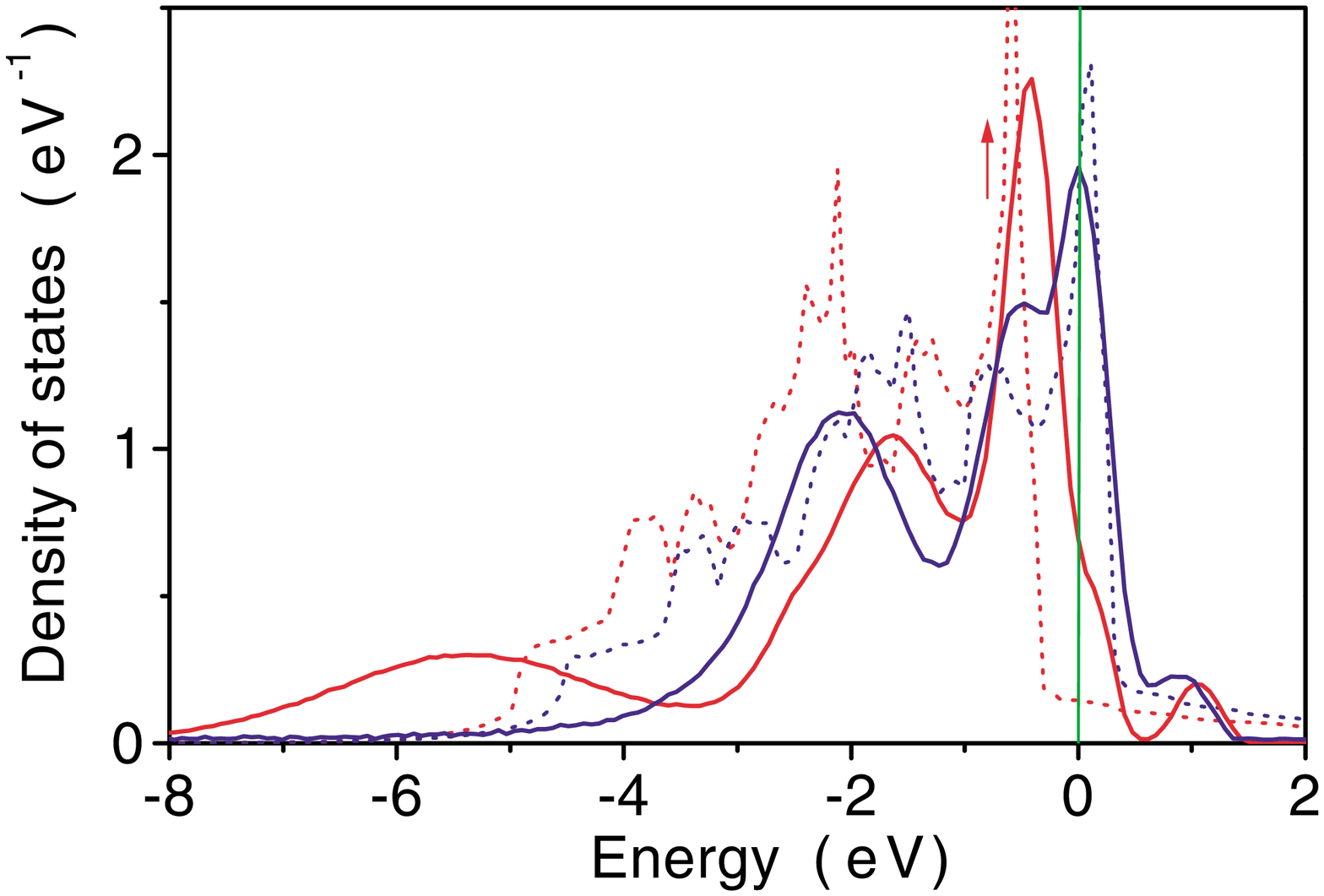}\\
\vspace{-5.3cm}

\hspace{6.11cm} \includegraphics[clip=true,width=7.8cm]{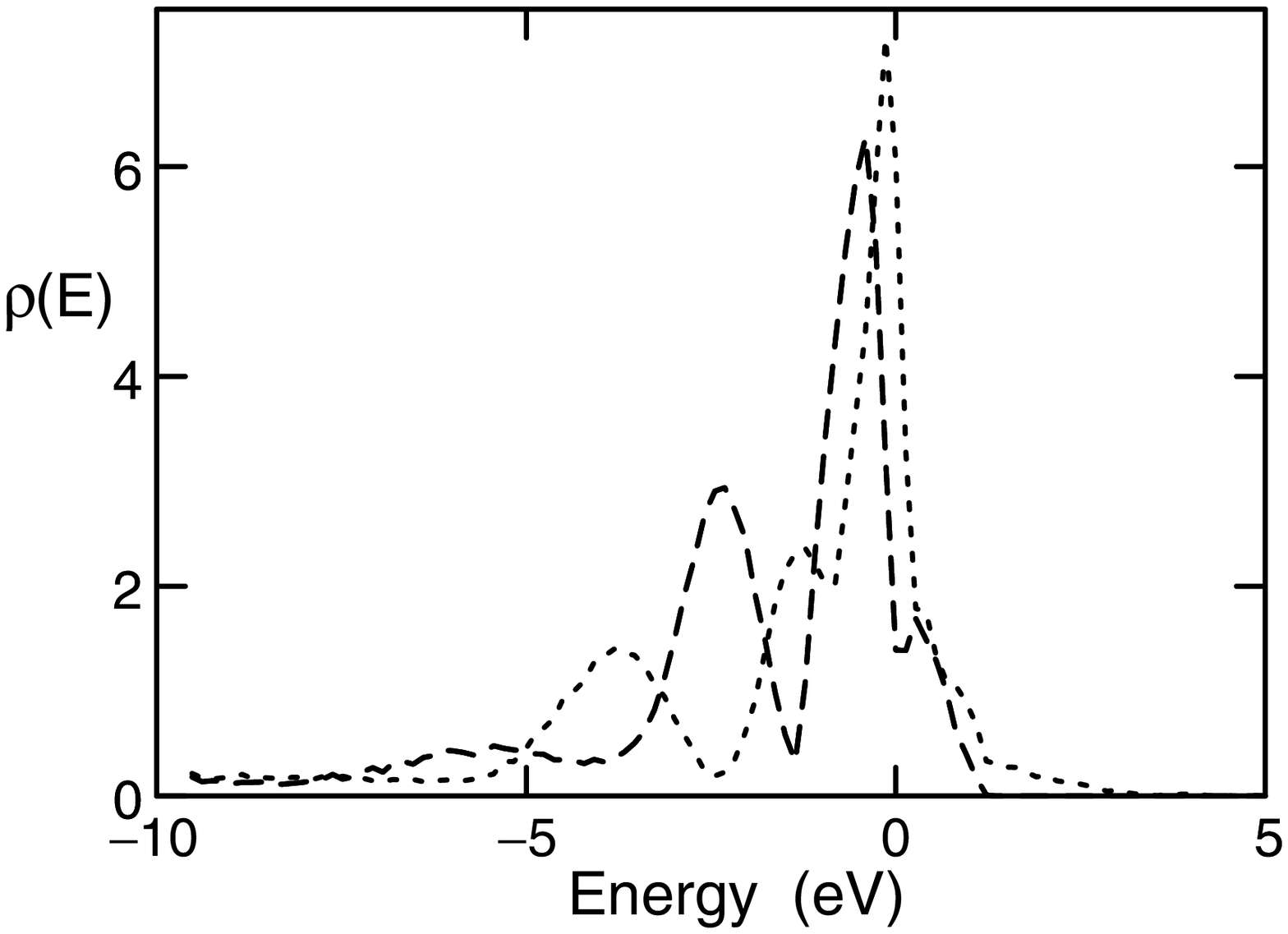}
}
\caption{\label{Fig:Ni}
Spectrum  ($\mathbf k$-integrated) of Ni [left: LDA+DMFT (solid lines), spinpolarized LDA (LSDA) (dotted lines); right: $GW$+DMFT].
The two lines represent the minority- and majority-spin spectrum respectively. 
At roughly -6eV, a satellite peak is clearly visible
in the majority-spin spectrum
(reproduced from \protect\cite{FeNi} and \protect\cite{Biermann03}, respectively).}
\end{figure}

Fig.\ \ref{Fig:Ni} (right panel) 
shows the $GW$+DMFT $\mathbf k$-integrated spectral function of Ni which is 
similar to LDA+DMFT results (left panel). Both approaches yield a satellite peak at $\approx -6\,$eV.

\section{All of that: {\em ab intito} D$\Gamma$A}
\label{Sec:DGA}



From the Hedin equations, it seems to be much more natural to connect
(i) the  $GW$ physics of screened exchange 
and (ii) strong, local correlations 
on the two-particle
level than on the one particle level as done in $GW$+DMFT.
In the Hedin equations, the natural starting point is 
the two-particle (particle-hole) irreducible
vertex. 
A generalization of DMFT to $n$-particle correlation functions
is the dynamical vertex approximation (D$\Gamma$A) \cite{Toschi06a}
 which approximates the  $n$-particle  fully irreducible\footnote{Fully irreducible means, cutting any two Green function lines does not separate the diagram 
into two parts. It is even more restrictive (less diagrams) than 
 the  particle-hole irreducible vertex (whose diagrams can be reducible e.g. in the particle-particle channel).} vertex $\Gamma_{\rm ir}$
 by  the corresponding local contribution of all Feynman diagrams. 
For $n=1$ the one-particle irreducible vertex is the self energy so that
D$\Gamma$A yields the DMFT. For $n=2$, we obtain non-local correlations
on all length scales and can calculate, e.g., the critical exponents of the Hubbard model
\cite{Rohringer}. 

\begin{figure}
    \centering\includegraphics[width=0.2\textwidth,clip=true,angle=270]{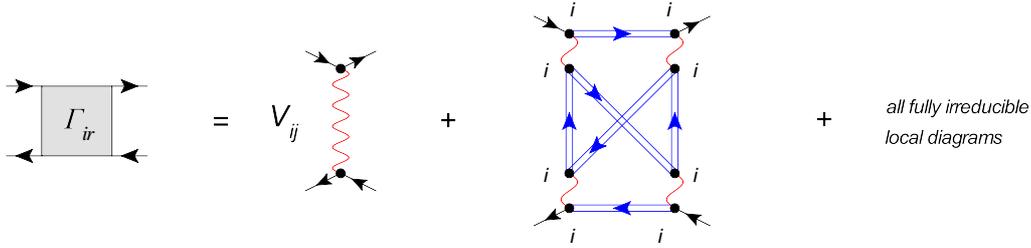}
    \caption{
In {\em ab initio} D$\Gamma$A we take as the fully irreducible
vertex the bare Coulomb interaction plus all local vertex corrections (only one such diagram is exemplarily shown) \cite{Held}.
    \label{Fig:DGA}}
  \end{figure}

Recently, some of us have proposed to use this D$\Gamma$A {\em ab initio}
for materials calculation \cite{Held}. The fully irreducible vertex $\Gamma_{\rm ir}$  is then
given by the bare Coulomb interaction, which possibly is non-local,
and all higher order local Feynman diagrams, see Fig.\ \ref{Fig:DGA}. From $\Gamma_{\rm ir}$ the  full (reducible) vertex
is calculated via the parqet equations \cite{Bickers}.
The calculation of the local part of  $\Gamma_{\rm ir}$ only requires us to calculate the
two-particle Green functions of a single-site Anderson impurity model, which is well doable even for realistic multi-orbital models. 
For the parquet equations on the other hand, there has 
been  some recent progress  \cite{Jarrell}.

As a simplified version of  {\em ab initio} D$\Gamma$A one can restrict
oneself to a subset of the three channels of the parquet equations,
as was done in  \cite{Toschi06a,Rohringer}.  In this case one has
to solve the  Bethe-Salpeter equation with the particle-hole
irreducible vertex  $\Gamma_{\rm ph}$
instead of the parquet equations with the fully irreducible vertex  $\Gamma_{\rm ir}$. That is, our approximation to the Hedin equations is to take 
the local $\Gamma^*_{\rm ph}$ (all Feynman diagrams given by the local Green function and interaction)
in the Hedin equation (\ref{Eq:BS}). In practice, one solves an Anderson impurity model numerically to calculate  $\Gamma^*_{\rm ph}$.

Full and simplified version of  {\em ab initio} D$\Gamma$A contain the
diagrams (and physics) of $GW$, DMFT as well as non-local correlations which
are responsible for (para-)magnons, (quantum) criticality  and ``all that''.

Support of the Austrian Science Fund (FWF) 
 through I597 (Austrian part of FOR 1346  with the Deutsche Forschungsgemeinschaft as lead agency) is
gratefully acknowledged.

\section*{Appendices}
\appendix
\section{Additional steps: equation of motion}
\label{Sec:appendix}

\begin{figure}[tb]
 \centering
 \includegraphics[width=0.19\textwidth,angle=270]{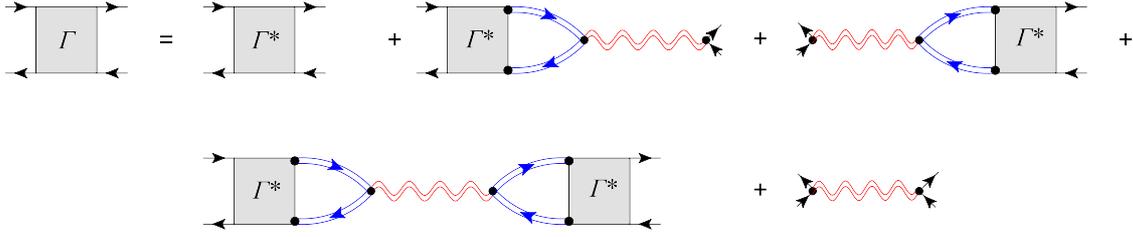}
 \caption{Relation between $\Gamma$ and $\Gamma^*$ in terms of Feynman diagrams.}
\label{Fig:gamma_gamma_star}
\end{figure}

In this appendix a detailed explanation is given how to derive the Hedin equation of motion for $\Sigma$, i.e., equation (\ref{Eq:HeisenbergHedin}), from
the standard equation of motion (\ref{Eq:SigmaVG}). In a first step $\Gamma$ is expressed in terms of $\Gamma^*$. In order to keep the notation simple,
the arguments of all functions are omitted and the functions are considered as operators. \\
The starting point of the calculations are the Bethe-Salpeter equations for $\Gamma$ and $\Gamma^*$:

\begin{equation}
\label{Eqapp:BetheSalpeter}
\begin{split}
&\Gamma=\Gamma_{\rm ph}+\Gamma\beta GG\Gamma_{\rm ph}\Longrightarrow \Gamma=\Gamma_{\rm ph}(1-\beta GG\Gamma_{\rm ph})^{-1}\\
&\Gamma^*=\Gamma^*_{\rm ph}+\Gamma^*_{\rm ph}\beta GG\Gamma^* \Longrightarrow \Gamma^*_{\rm ph}=\Gamma^*(1+\beta GG\Gamma^*)^{-1}.\\
\end{split}
\end{equation}\
Using  equation (\ref{Eq:Gammastardef}), i.e., $\Gamma_{\rm ph}=\Gamma^*_{\rm ph}+V$, one gets:
\begin{equation}
\label{Eqapp:Gamma1}
\Gamma=\biggl(\Gamma^*(1+\beta GG\Gamma^*)^{-1}+V\biggr)\frac{1}{1-\beta GG\biggl(\Gamma^*(1+\beta GG\Gamma^*)^{-1}+V\biggr)}
\end{equation}
Multiplying this equation with $1=(1+\beta GG\Gamma^*)^{-1}(1+\beta GG\Gamma^*)$ and using the standard relations for operators, $A^{-1}B^{-1}=(BA)^{-1}$
leads to:
\begin{equation}
\Gamma=\biggl(\Gamma^*(1+\beta GG\Gamma^*)^{-1}+V\biggr)\underset{1}{\underbrace{V^{-1}V}}\biggl(1-(\underset{P}{\underbrace{\beta GG+\beta GG\Gamma^*\beta GG}})V\biggr)^{-1}\biggl(1+\beta GG\Gamma^*\biggr),
\end{equation}
where $1=V^{-1}V$ was inserted and the definition for the polarization operator, equation (\ref{Eq:HedinP}). Now one can use the second Hedin equation (\ref{Eq:W}), which can be rewritten as $W=V(1-PV)^{-1}$.  Inserting this relation into the equation for $\Gamma$, one arrives at the following result:
\begin{equation}
\Gamma=\biggl(\Gamma^*(1+\beta GG\Gamma^*)^{-1}\biggr)V^{-1}W\biggl(1+\beta GG\Gamma^*\biggr)+W\biggl(1+\beta GG\Gamma^*\biggr).
\end{equation}
Another formulation of the second Hedin equation gives $V^{-1}=W^{-1}+P=W^{-1}+(1+\beta GG \Gamma^* )\beta GG$. Replacing $V^{-1}$ by this expression gives:
\begin{equation}
\begin{split}
 \label{Eqapp:Gammafinal}
\Gamma&=\Gamma^*+\Gamma^*\beta GGW(1+\beta GG\Gamma^*)+W(1+\beta GG\Gamma^*)\\
&=\Gamma^*+\Gamma^* \beta GGW+\Gamma^* \beta GGW \beta GG \Gamma^*+ W \beta GG\Gamma^*+W\; .
\end{split}
\end{equation}
This equation shows how the full $\Gamma$ is related to the $\Gamma^*$.  Diagrammatically this relation is shown in Fig. \ref{Fig:gamma_gamma_star}.

In the next step $\Gamma$ as given in equation (\ref{Eqapp:Gammafinal}) is inserted into equation (\ref{Eq:SigmaVG}), yielding 
\begin{equation}
\begin{split}
\label{Eqapp:Sigma}
\Sigma&=-GV\beta GG \Gamma-GV\\&=-GV(\underset{P}{\underbrace{\beta GG+\beta GG\Gamma^*\beta GG}})W-GV\bigl((\underset{P}{\underbrace{\beta GG +\beta GG\Gamma^*\beta GG}})W+1\bigr)\beta GG \Gamma^*-GV \; .
\end{split}
\end{equation}
From the second Hedin equation it follows that $VPW=W-V$. Inserting this relation into Equation  (\ref{Eqapp:Sigma}) yields:
\begin{equation}
\begin{split}
\Sigma&=-G(W-V)-G(W-V)\beta GG \Gamma^*-GV\beta GG \Gamma^*-GV=\\
&=-GW-GW\beta GG \Gamma^*,
\end{split}
\end{equation}
which is exactly equation (\ref{Eq:HeisenbergHedin}).

Let us also,
at this point, mention that $\Gamma$ should satisfy an important relation:
\begin{equation} 
\Gamma(11';22')=-\Gamma(12';21')
\end{equation}
This relation is known as {\bf crossing symmetry} (see e.g. \cite{Bickers}, equation 7.5) and is simply a consequence of the Pauli-principle: Exchanging two identical fermions leads to a $-$ sign in the wave function. The screened interaction $W$, however, does not fulfill this crossing symmetry. Therefore, setting $\Gamma^*=0$, as it is done in the $GW$-approximation leads to  $\Gamma=W$ (see Fig.\ \ref{Fig:gamma_gamma_star}) which violates this crossing symmetry, i.e. it violates the Pauli-principle. \\


\end{document}